\documentclass[twocolumn]{jpsj3}

\usepackage{color}
\usepackage{bm}

\title{%
Time-Dependent Ginzburg--Landau Equation and Boltzmann Transport Equation
for Charge-Density-Wave Conductors}

\author{%
Yositake Takane$^1$, Masahiko Hayashi$^2$, and Hiromichi Ebisawa$^3$}

\inst{%
$^1$Department of Quantum Matter,
Graduate School of Advanced Sciences of Matter,\\
Hiroshima University, Higashihiroshima, Hiroshima 739-8530, Japan\\
$^2$Faculty of Education and Human Studies, Akita University,
Akita 010-8502, Japan\\
$^3$Graduate School of Information Sciences, Tohoku University,
Sendai 980-8579, Japan
}

\recdate{ \hspace{50mm} }

\abst{%
The time-dependent Ginzburg--Landau equation and the Boltzmann transport
equation for charge-density-wave (CDW) conductors
are derived from a microscopic one-dimensional model by applying
the Keldysh Green's function approach under a quasiclassical approximation.
The effects of an external electric field and impurity pinning of the CDW
are fully taken into account without relying on a phenomenological argument.
These equations simultaneously describe the spatiotemporal dynamics of
both the CDW and quasiparticles; thus, they serve as a starting point
to develop a general framework to analyze various nonequilibrium phenomena,
such as current conversion between the CDW condensate and quasiparticles,
in realistic CDW conductors.
It is shown that, in typical situations, the equations correctly
describe the nonlinear behavior of electric conductivity in a simpler manner.
}

\begin{document}
\sloppy
\maketitle

\section{Introduction}

The advances in experimental technology have enabled us to closely study
the dynamical properties of
charge-density-wave (CDW) conductors.~\cite{monceau}
Several unique behaviors of CDW conductors have been observed
at nonequilibrium.~\cite{mantel,markovic,vanderzant,ayari,oneill,slot,rai,
isakovic,matsuura1,matsuura2,zhang,sinchenko}
An interesting example is the current conversion between the CDW condensate
and quasiparticles mediated by phase slips,
which has been the subject of intensive experimental studies.~\cite
{maher1,maher2,adelman,lemay,requardt}
Although plausible models have been proposed for this problem,~\cite{ong1,ong2,
brazovskii1,brazovskii2,ramakrishna,duan,maki,hatakenaka,
brazovskii3,takane1,takane2,takane3,takane4}
a reliable picture of the current conversion process has not been established.
To fully understand the nonequilibrium behaviors of CDW conductors,
we need to describe the dynamics of both the CDW and quasiparticles
including the effects of an external electric field and
pinning of the CDW due to impurities.
In principle, the Keldysh Green's function approach provides
a concrete theoretical framework that fulfills the requirement
mentioned above.~\cite{artemenko1,artemenko2,artemenko3,eckern1,
artemenko4,artemenko5,takane5}
However, this is not easy to handle,
even under a quasiclassical approximation.

To make the framework more tractable, a practical way is to reduce
it to a set comprising the time-dependent Ginzburg--Landau (TDGL) equation
for the CDW order parameter and the Boltzmann transport (BT) equation
for charge and current densities
by expanding it with respect to the CDW order parameter.
These equations are written in the form of differential equations, and hence
are easy to handle, although their application may be restricted to the regime
where the magnitude of the CDW order parameter is sufficiently small.
Such a framework based on the TDGL and BT equations has been fully developed
and is widely used in the field of superconductivity,~\cite{kopnin}
while it remains inadequate in the field of CDW conductors.
Previously, the TDGL equation for CDW conductors has been proposed and
used to analyze dynamical features of the CDW.~\cite{gorkov1,gorkov2,batistic}
However, the proposed TDGL equation describes only the CDW degrees of freedom
and the role of quasiparticles is neglected.
The BT equation for CDW conductors has been introduced
by several authors.~\cite{eckern1,su,ishikaw}
However, its application is restricted to the case without pinning of the CDW.

In this paper, we derive the TDGL and BT equations for CDW conductors
starting from a microscopic one-dimensional model.
The resulting equations fully describe the dynamics of
both the CDW and quasiparticles including
the effects of an external electric field and impurity pinning
in one-dimensional situations,
and serve as a starting point to develop a general framework to analyze
nonequilibrium phenomena in realistic CDW conductors.
They are derived by applying
the Keldysh Green's function approach~\cite{keldysh,rammer}
under a quasiclassical approximation~\cite{eilenberger,larkin}
without relying on a phenomenological argument.
In addition to the ordinary assumption that the magnitude of
the CDW order parameter $\Delta(x,t)$ is much smaller than the temperature $T$,
we assume that the nonequilibrium distribution of quasiparticles
is described by the Fermi-Dirac function with a space- and time-dependent
chemical potential $\mu_{\pm}(x,t)$, where $\mu_{+}(x,t)$ [$\mu_{-}(x,t)$]
denotes the chemical potential for the right-going (left-going) quasiparticles
[see Eq.~(\ref{eq:f+-})].
The dynamical variables in our problem are
$\Delta(x,t)$, $\mu_{+}(x,t)$, and $\mu_{-}(x,t)$,
by which the dynamics of the CDW and quasiparticles are described
including the effects of an external electric field and impurity pinning.

In the next section, we present a one-dimensional model for CDW conductors
and introduce the Keldysh Green's function with its equation
of motion under a quasiclassical approximation.
In Sect.~3, we derive the TDGL and BT equations by using
the standard procedure of the Keldysh Green's function approach
developed in the field of superconductivity.
In Sect.~4, we analyze the nonlinear behavior of electric conductivity
in typical situations within the framework of the TDGL and BT equations.
It is shown that the previously reported result is correctly reproduced
in a simpler manner.
The last section is devoted to a short summary.
Preliminary results of this work have been briefly reported
in Refs.~\citen{hayashi1} and \citen{hayashi2}.
We ignore the spin degree of freedom
and set $\hbar = k_{\rm B} = 1$ throughout this paper.

\section{Model and Formulation}

To present a one-dimensional model for CDW conductors,
it is convenient to decompose the electron field operator
$\psi(x)$ into the right-going and left-going components:
$\psi(x) = e^{ik_{\rm F}x}\psi_{+}(x) + e^{-ik_{\rm F}x}\psi_{-}(x)$
with $k_{\rm F}$ being the Fermi wave number.
The model Hamiltonian is given by $H = H_{0}+H_{\rm imp}$ with
\begin{align}
 & H_{0} = \int dx
     \Big\{  \psi_{+}^{\dagger}(x)d_{+}(x,t)\psi_{+}(x)
           + \psi_{-}^{\dagger}(x)d_{-}(x,t)\psi_{-}(x)
       \nonumber \\
 & \hspace{20mm}
      + g u(x,t) e^{-iQx} \psi_{+}^{\dagger}(x)\psi_{-}(x)
      + {\rm h.c.}
     \Big\} ,
       \\
 & H_{\rm imp} = \int dx
     \Big\{  \psi_{+}^{\dagger}(x) V_{\rm imp}(x) \psi_{+}(x)
           + \psi_{-}^{\dagger}(x) V_{\rm imp}(x) \psi_{-}(x)
        \nonumber \\
 & \hspace{20mm}
           + \psi_{+}^{\dagger}(x)
             e^{-iQx}V_{\rm imp}(x) \psi_{-}(x)
           + {\rm h.c.}
     \Big\} ,
\end{align}
where
\begin{align}
   d_{\pm}(x,t)
   = \mp iv_{\rm F}\left[\partial_{x}+ ieA(x,t)\right] + \Phi(x,t) ,
\end{align}
$Q=2k_{\rm F}$, $g$ is the coupling constant between
electrons and phonons, and $u(x,t)$ is the lattice displacement,
which is directly connected with phonon degrees of freedom.
Here, $\Phi(x,t) = -e \phi(x,t)$ with $\phi$ being the scalar potential.
An electric field is expressed in terms of $\Phi$ and $A$
in a gauge-invariant manner [see Eq.~(\ref{eq:def-E})].
We assume that the impurity potential $V_{\rm imp}$ is given by
\begin{align}
     \label{eq:def-V_imp}
  V_{\rm imp}(x) = \sum_{i}v(x-x_{i}) ,
\end{align}
where $x_{i}$ denotes the location of the $i$th impurity.

In the previous theoretical treatment based on the Keldysh Green's function
approach,~\cite{artemenko1,artemenko2,artemenko3,eckern1,artemenko4,artemenko5}
the influence of $H_{\rm imp}$ is taken into account
as a correction to the self-energy
under the averaging over impurity configurations (i.e., disorder average).
As a first-order correction to the self-energy vanishes
under the disorder average, a nonvanishing contribution originates from
a second-order correction, which describes quasiparticle scattering.
An apparent drawback of this treatment is that impurity pinning
completely disappears in the resulting framework.
This is simply because the translational invariance of the system
is restored under the disorder average
in spite of the fact that a spatial inhomogeneity is indispensable for
the pinning of CDWs.

The pinning of CDWs is induced by backward scattering between right-going
and left-going electrons;~\cite{fukuyama1,fukuyama2,tutto,eckern2}
thus, the corresponding contribution should arise
from the third and fourth terms in $H_{\rm imp}$ containing
\begin{align}
        \label{eq:def-zeta}
   \zeta(x) \equiv e^{-iQx}V_{\rm imp}(x)
\end{align}
or its complex conjugate.
We capture the effect of impurity pinning
by explicitly incorporating it into our consideration.
Note that $\zeta(x)$ vanishes under the disorder average;
thus, we keep in mind that it is defined under a given impurity configuration.
Since $\zeta(x)$ plays a similar role to $g u(x,t)e^{-iQx}$ in $H_{0}$,
it is natural to define the order parameter $\Delta(x,t)$ for the CDW
by adding $\zeta(x)$ to $g \langle u(x,t) \rangle e^{-iQx}$ as
\begin{align}
  \Delta(x,t) \equiv g \langle u(x,t) \rangle e^{-iQx} + \zeta(x) ,
\end{align}
where $\langle \cdots \rangle$ denotes the thermal average.~\cite{comment}
Note that $\zeta(x)$ in the definition of $\Delta(x,t)$ can be regarded as
a first-order correction to the self-energy.
We hereafter consider the following effective Hamiltonian:
\begin{align}
      \label{eq:H-eff}
 H_{\rm eff}
 & = \int dx
     \Big\{  \psi_{+}^{\dagger}(x)d_{+}(x,t)\psi_{+}(x)
           + \psi_{-}^{\dagger}(x)d_{-}(x,t)\psi_{-}(x)
          \nonumber \\
 & \hspace{5mm}
         + \Delta(x,t) \psi_{+}^{\dagger}(x)\psi_{-}(x)
         + {\rm h.c.}
     \Big\} ,
\end{align}
instead of $H_{0}$.
Although $\zeta(x)$ seemingly disappears in $H_{\rm eff}$,
it certainly describes the impurity pinning via the self-consistency
equation for $\Delta(x,t)$ [see Eq.~(\ref{eq:gap_eq0})].
In the argument given below, the effect of quasiparticle scattering
is taken into account as an ordinary second-order correction
to the self-energy [see Eq.~(\ref{eq:sigma-imp})].

To describe the dynamics of the CDW and quasiparticles governed by
$H_{\rm eff}$ with $H_{\rm imp}$,
we introduce the Keldysh Green's function consisting of
the Keldysh, retarded, and advanced components:~\cite{keldysh}
\begin{align}
  & G_{\alpha \beta}^{K}(x,t;x',t')
    = - i
       \nonumber \\
  & \hspace{10mm}
          \times
          \Big\langle \psi_{\alpha}(x,t)\psi_{\beta}^{\dagger}(x',t')
          - \psi_{\beta}^{\dagger}(x',t')\psi_{\alpha}(x,t) \Big\rangle ,
          \\
  & G_{\alpha \beta}^{R}(x,t;x',t')
    = - i \Theta(t-t')
       \nonumber \\
  & \hspace{10mm}
          \times
          \Big\langle \psi_{\alpha}(x,t)\psi_{\beta}^{\dagger}(x',t')
          + \psi_{\beta}^{\dagger}(x',t')\psi_{\alpha}(x,t) \Big\rangle ,
          \\
  & G_{\alpha \beta}^{A}(x,t;x',t')
    = + i \Theta(t'-t)
       \nonumber \\
  & \hspace{10mm}
          \times
          \Big\langle \psi_{\alpha}(x,t)\psi_{\beta}^{\dagger}(x',t')
          + \psi_{\beta}^{\dagger}(x',t')\psi_{\alpha}(x,t) \Big\rangle ,
\end{align}
where $\alpha, \beta (= \pm)$ specify
the right-going and left-going components,
and $\Theta(t)$ is Heaviside's step function.
In accordance with the effective Hamiltonian, the matrix Green's functions
$\hat{G}^{X}(x,t;x',t')$ ($X = K, R, A$), defined by
\begin{equation}
 \hat{G}^{X}(x,t;x',t')
    = \left[ \begin{array}{cc}
               G_{++}^{X} & G_{+-}^{X} \\
               G_{-+}^{X} & G_{--}^{X}
             \end{array}
      \right] ,
\end{equation}
satisfy
\begin{align}
       \label{eq:eq-of-motion_G^K}
 \hat{\mathcal D}(x,t)\hat{G}^{K}(x,t;x',t')
 & =  \left\{ \hat{\Sigma}_{\ast}^{R}
              \otimes\hat{G}^{K} \right\}(x,t;x',t')
          \nonumber \\
 & \hspace{-15mm}
    + \left\{ \hat{\Sigma}_{\ast}^{K}
              \otimes\hat{G}^{A} \right\}(x,t;x',t') ,
           \\
       \label{eq:eq-of-motion_G^R,A}
 \hat{\mathcal D}(x,t)\hat{G}^{R,A}(x,t;x',t')
 & = \hat{\sigma}_{0}\delta(x-x')\delta(t-t')
          \nonumber \\
 & \hspace{-15mm}
     + \left\{ \hat{\Sigma}_{\ast}^{R,A}
               \otimes\hat{G}^{R,A} \right\}(x,t;x',t') ,
\end{align}
where 
\begin{align}
   \hat{\mathcal D}(x,t)
 & = \hat{\sigma}_{0} \left[i\partial_{t}- \Phi(x,t)\right]
     + i\hat{\sigma}_{z}v_{\rm F}\left[\partial_{x}+ ieA(x,t)\right]
               \nonumber \\
 & \hspace{10mm}
      - \hat{\Delta}_{\ast}(x,t)
\end{align}
with
\begin{align}
    \hat{\Delta}_{\ast}(x,t)
    = \left[ \begin{array}{cc}
                0 & \Delta \\
                \Delta^{*} & 0
             \end{array}           \right] ,
\end{align}
and $\left\{ A \otimes B \right\}(x,t;x',t')$ denotes
\begin{align}
 & \left\{ A \otimes B \right\}(x,t;x',t')
      \nonumber \\
 & \hspace{7mm}
   = \int dx_{1} \int dt_{1}
     A(x,t;x_{1},t_{1})B(x_{1},t_{1};x'.t') .
\end{align}
Here and hereafter, we use $\hat{\sigma}_{0}$ and $\hat{\sigma}_{i}$
($i = x,y,z$) to denote the $2 \times 2$ unit matrix and
the $i$-component of the Pauli matrix, respectively.
The self-energy $\hat{\Sigma}_{\ast}^{X}$
describing the impurity scattering of quasiparticles is given by
\begin{align}
      \label{eq:sigma-imp}
  \hat{\Sigma}_{\ast}^{X}(x,t;x',t')
  = \left\langle
    \hat{V}_{\rm imp}(x)\hat{G}^{X}(x,t;x',t')\hat{V}_{\rm imp}(x')
    \right\rangle_{\rm imp} ,
\end{align}
where $\langle\cdots\rangle_{\rm imp}$ denotes the disorder average and
\begin{equation}
  \hat{V}_{\rm imp}(x)
    = V_{\rm imp}(x)
      \left[ \begin{array}{cc}
               1 & e^{-iQx} \\
               e^{iQx} & 1
             \end{array}
      \right] .
\end{equation}
The equations for the matrix Green's functions are supplemented by
the self-consistency equation for the order parameter:
\begin{align}
     \label{eq:gap_eq0}
   \Delta(x,t)
     = 4 \pi i v_{\rm F} \lambda
       \left. G^{K}_{+-}(xt,x't') \right|_{x' \to x, t' \to t}
     + \zeta(x) ,
\end{align}
where $\lambda$ is a dimensionless coupling constant.
The derivation of Eq.~(\ref{eq:gap_eq0}) is outlined in Appendix~A.
We see below that $\zeta(x)$ in Eq.~(\ref{eq:gap_eq0})
gives rise to impurity pinning.

With the Fourier transform of $\hat{G}^{X}$,
\begin{align}
 & \hat{G}^{X}(x,k;t,\epsilon) \equiv
   \int dy e^{-iky}
   \int d\tau e^{i\epsilon\tau}
        \nonumber \\
 & \hspace{5mm}
   \times
   \hat{G}^{X} \left(x+\frac{y}{2},t+\frac{\tau}{2};
                     x-\frac{y}{2},t-\frac{\tau}{2}\right) ,
\end{align}
we introduce the quasiclassical Green's function:~\cite{eilenberger}
\begin{equation}
 \hat{g}^{X}(x;t,\epsilon) =
   \hat{\sigma}_{z} \frac{iv_{\rm F}}{\pi}
   \int dk \hat{G}^{X}(x,k;t,\epsilon) ,
\end{equation}
where a diverging contribution must be subtracted.
The self-consistency equation is rewritten as
\begin{align}
      \label{eq:reduced-gap_eq}
   \Delta(x,t)
   = \lambda\int d\epsilon
     g^{K}_{+-}(x;t,\epsilon) + \zeta(x) .
\end{align}
The charge density $\rho(x,t)$ and the current density $j(x,t)$
are expressed as
\begin{align}
      \label{eq:charge_density}
  \rho(x,t)
 & = \frac{e}{\pi v_{\rm F}}
      \left( \frac{1}{8}\int d\epsilon \,
             {\rm tr}\left\{ \hat{\sigma}_{z}\hat{g}^{K}(x;t,\epsilon)
                     \right\} + \Phi(x,t)
      \right) ,
          \\
      \label{eq:current_density}
  j(x,t)
 & = \frac{e}{8\pi}
     \int d\epsilon \,
             {\rm tr}\left\{ \hat{g}^{K}(x;t,\epsilon)\right\} .
\end{align}
Within a quasiclassical approximation, we can derive a kinetic equation for
$\hat{g}^{X}(x;t,\epsilon)$ from Eqs.~(\ref{eq:eq-of-motion_G^K})
and (\ref{eq:eq-of-motion_G^R,A}):
\begin{align}
    \label{eq:kin-G}
&   (\epsilon - \Phi)\left[\hat{\sigma}_{z}, \hat{g}^{X} \right]_{-}
  + \frac{i}{2}\left[\hat{\sigma}_{z}, \dot{\hat{g}}^{X} \right]_{+}
  + i v_{\rm F}\partial_{x}\hat{g}^{X}
  + \left[\hat{\Delta}, \hat{g}^{X} \right]_{-}
        \nonumber \\
& \hspace{5mm}
  + i ev_{\rm F}\dot{A}\partial_{\epsilon}\hat{g}^{X}
  + \frac{i}{2}\dot{\Phi}\left[\hat{\sigma}_{z},
                     \partial_{\epsilon}\hat{g}^{X} \right]_{+}
  - \frac{i}{2}\left[\dot{\hat{\Delta}},
                     \partial_{\epsilon}\hat{g}^{X} \right]_{+}
       \nonumber \\
& \hspace{5mm}
  - \frac{1}{8}\left[\ddot{\hat{\Delta}},
                     \partial_{\epsilon}^{2}\hat{g}^{X} \right]_{-}
  - \left[\hat{\Sigma}^{X}, \hat{g}^{X} \right]_{-} = 0 ,
\end{align}
where $\hat{\Delta} \equiv - \hat{\Delta}_{\ast}\hat{\sigma}_{z}$.
The self-energy
$\hat{\Sigma}^{X}\equiv \hat{\Sigma}_{\ast}^{X}\hat{\sigma}_{z}$
is expressed as~\cite{artemenko1}
\begin{align}
      \label{eq:self-E_X}
   \hat{\Sigma}^{X}(x;t,\epsilon)
 = - \frac{i}{2}
     \left[  \nu_{1}\hat{\sigma}_{z}\hat{g}^{X}\hat{\sigma}_{z}
           - \frac{\nu_{2}}{2}
             \left( \hat{\sigma}_{x}\hat{g}^{X}\hat{\sigma}_{x}
                  + \hat{\sigma}_{y}\hat{g}^{X}\hat{\sigma}_{y}
             \right)
     \right] ,
\end{align}
where $\nu_{1}$ and $\nu_{2}$ respectively characterize the strength of
forward and backward scattering.
They are given by
$\nu_{1} = n_{\rm imp}\left|v(0)\right|^{2}/v_{\rm F}$ and
$\nu_{2} = n_{\rm imp}\left|v(Q)\right|^{2}/v_{\rm F}$,
where $n_{\rm imp}$ and $v(q)$ are respectively the density of impurities
and the Fourier transform of the impurity potential $v(x)$.
For later convenience, we define the elastic relaxation time $\tau$ and
the transport relaxation time $\tau_{\rm tr}$ as
\begin{align}
  \frac{1}{2\tau} & = \nu_{1} + \frac{\nu_{2}}{2} ,
       \\
      \label{eq:def-tau_tr}
  \frac{1}{2\tau_{\rm tr}} & = \nu_{2} .
\end{align}
The self-energy term in Eq.~(\ref{eq:kin-G}) for the Keldysh component reads
\begin{align}
       \label{eq:self-E_K}
  \left[\hat{\Sigma}^{K}, \hat{g}^{K} \right]_{-}
    = \hat{\Sigma}^{R}\hat{g}^{K}
      + \hat{\Sigma}^{K}\hat{g}^{A}
      - \hat{g}^{R}\hat{\Sigma}^{K}
      - \hat{g}^{K}\hat{\Sigma}^{A} .
\end{align}
We adopt the following approximate expression
for $\hat{g}^{K}$:~\cite{larkin,rammer}
\begin{align}
     \label{eq:gK-expression}
   \hat{g}^{K}(x;t,\epsilon)
 & =  \hat{g}^{R}\hat{n} - \hat{n}\hat{g}^{A}
    -\frac{i}{2}
     \left[ \partial_{t}\hat{g}^{R}
            \partial_{\epsilon}\hat{n}
          + \partial_{\epsilon}\hat{n}
            \partial_{t}\hat{g}^{A} \right]
         \nonumber \\
 &  \hspace{-13mm}
    +\frac{i}{2}
     \left[ \partial_{\epsilon}\hat{g}^{R}
            \partial_{t}\hat{n}
          + \partial_{t}\hat{n}
            \partial_{\epsilon}\hat{g}^{A} \right]
    -\frac{1}{8}
     \left[ \partial_{t}^{2}\hat{g}^{R}
            \partial_{\epsilon}^{2}\hat{n}
          - \partial_{\epsilon}^{2}\hat{n}
            \partial_{t}^{2}\hat{g}^{A} \right]
\end{align}
with
\begin{align}
  \hat{n}(x;t,\epsilon)
 = \left[ \begin{array}{cc}
             1 - 2 f_{+}(x;t,\epsilon) & 0 \\
             0 & 1 - 2 f_{-}(x;t,\epsilon)
          \end{array}
   \right] ,
\end{align}
where $f_{+}$ ($f_{-}$) denotes the distribution function for
right-going (left-going) quasiparticles.
Now we employ the assumption that $f_{+}$ and $f_{-}$ are expressed as
\begin{align}
       \label{eq:f+-}
  f_{\pm}(x,t,\epsilon)
    & = f_{\rm FD}\left( \epsilon-\Phi(x,t)-\mu_{\pm}(x,t) \right) ,
\end{align}
where $f_{\rm FD}(\epsilon)$ is the Fermi-Dirac function
and $\mu_{+}$ ($\mu_{-}$) is the chemical potential
for right-going (left-going) quasiparticles.

\section{Derivation of the TDGL and BT Equations}

To derive the TDGL and BT equations,
we solve Eq.~(\ref{eq:kin-G}) for $\hat{g}^{R}$ and $\hat{g}^{A}$.
With the explicit representation of
\begin{align}
  \hat{g}^{R}(x;t,\epsilon)
 = \left[ \begin{array}{cc}
             g^{R}(x;t,\epsilon) & f^{R}(x;t,\epsilon) \\
             \bar{f}^{R}(x;t,\epsilon) & \bar{g}^{R}(x;t,\epsilon)
          \end{array}
   \right] ,
\end{align}
Eq.~(\ref{eq:kin-G}) for $\hat{g}^{R}$ is decomposed into the four equations
given in Appendix~B.
Assuming that $|\Delta| \ll T$, we approximately solve them
with the help of the normalization condition in the static limit:
\begin{align}
 (g^{X})^{2}+f^{X}\bar{f}^{X} = (\bar{g}^{X})^{2}+f^{X}\bar{f}^{X} = 1 ,
\end{align}
where $X=R$ or $A$.
We find that the matrix elements are given by
\begin{align}
     \label{eq:g^R-f}
  g^{R}(x;t,\epsilon)
  & = 1 + \frac{|\Delta|^{2}}{2(\epsilon-\Phi+\frac{i}{2\tau})^{2}}
        \nonumber \\
  & \hspace{-10mm}
      - \frac{iv_{\rm F}}{4(\epsilon-\Phi+\frac{i}{2\tau})^{3}}
        \left( \Delta^{*}\partial_{x}\Delta
               - \Delta\partial_{x}\Delta^{*} \right)
        \nonumber \\
  & \hspace{-10mm}
      + \frac{i}{4(\epsilon-\Phi+\frac{i}{2\tau})^{3}}
        \left( \Delta^{*}\partial_{t}\Delta
               - \Delta\partial_{t}\Delta^{*} \right)  ,
           \\
  \bar{g}^{R}(x;t,\epsilon)
  & = - 1 - \frac{|\Delta|^{2}}{2(\epsilon-\Phi+\frac{i}{2\tau})^{2}}
        \nonumber \\
  & \hspace{-10mm}
      + \frac{iv_{\rm F}}{4(\epsilon-\Phi+\frac{i}{2\tau})^{3}}
        \left( \Delta^{*}\partial_{x}\Delta
               - \Delta\partial_{x}\Delta^{*} \right)
        \nonumber \\
  & \hspace{-10mm}
      + \frac{i}{4(\epsilon-\Phi+\frac{i}{2\tau})^{3}}
        \left( \Delta^{*}\partial_{t}\Delta
               - \Delta\partial_{t}\Delta^{*} \right)  ,
           \\
  f^{R}(x;t,\epsilon)
  & = \frac{\Delta}{\epsilon-\Phi+\frac{i}{2\tau}}
      - \frac{iv_{\rm F}}{2(\epsilon-\Phi+\frac{i}{2\tau})^{2}}
        \partial_{x}\Delta
        \nonumber \\
  & \hspace{-10mm}
      - \frac{\epsilon-\Phi}{2(\epsilon-\Phi+\frac{i}{2\tau})^{4}}
        \Delta |\Delta|^{2}
        \nonumber \\
  & \hspace{-10mm}
      - \frac{iv_{\rm F}eE}{2(\epsilon-\Phi+\frac{i}{2\tau})^{3}}
        \Delta
      - \frac{v_{\rm F}^{2}}{4(\epsilon-\Phi+\frac{i}{2\tau})^{3}}
        \partial_{x}^{2}\Delta ,
           \\
     \label{eq:bar_f^R-f}
  \bar{f}^{R}(x;t,\epsilon)
  & = - \frac{\Delta^{*}}{\epsilon-\Phi+\frac{i}{2\tau}}
      - \frac{iv_{\rm F}}{2(\epsilon-\Phi+\frac{i}{2\tau})^{2}}
        \partial_{x}\Delta^{*}
        \nonumber \\
  & \hspace{-10mm}
      - \frac{\epsilon-\Phi}{2(\epsilon-\Phi+\frac{i}{2\tau})^{4}}
        \Delta^{*} |\Delta|^{2}
        \nonumber \\
  & \hspace{-10mm}
      - \frac{iv_{\rm F}eE}{2(\epsilon-\Phi+\frac{i}{2\tau})^{3}}
        \Delta^{*}
      + \frac{v_{\rm F}^{2}}{4(\epsilon-\Phi+\frac{i}{2\tau})^{3}}
        \partial_{x}^{2}\Delta^{*} ,
\end{align}
where $E$ denotes the electric field defined by
\begin{align}
         \label{eq:def-E}
 E(x,t) = \frac{1}{e}\partial_{x}\Phi(x,t)-\partial_{t}A(x,t) .
\end{align}
The advanced function $\hat{g}^{A}$ is obtained via the relation
\begin{align}
  \hat{g}^{A}(x;t,\epsilon) = -\hat{g}^{R}(x;t,\epsilon)
  |_{\frac{i}{2\tau} \to -\frac{i}{2\tau}} .
\end{align}

Approximating $\hat{g}^{K}$ by the first term of the gradient expansion
of Eq.~(\ref{eq:gK-expression}),
namely, $\hat{g}^{K} = \hat{g}^{R}\hat{n} - \hat{n}\hat{g}^{A}$,
and substituting this with Eqs.~(\ref{eq:g^R-f})--(\ref{eq:bar_f^R-f})
into Eqs.~(\ref{eq:charge_density}) and (\ref{eq:current_density}),
we readily find that the charge and current densities are given by
\begin{align}
      \label{eq:rho-f}
  \rho(x,t)
    & = -\frac{e}{2\pi v_{\rm F}}
         \left(1-\frac{7\zeta(3)}{4\pi^{2}T^{2}}|\Delta|^{2} \right)
         (\mu_{+}+\mu_{-})
        \nonumber \\
  & \hspace{0mm}
        + ie \frac{7\zeta(3)}{16\pi^{3}T^{2}}
          \left( \Delta^{*}\partial_{x}\Delta - \Delta\partial_{x}\Delta^{*}
          \right) ,
           \\
      \label{eq:j-f}
  j(x,t)
    & = -\frac{e}{2\pi}
         \left(1-\frac{7\zeta(3)}{4\pi^{2}T^{2}}|\Delta|^{2} \right)
         (\mu_{+}-\mu_{-})
        \nonumber \\
  & \hspace{0mm}
       - ie \frac{7\zeta(3)}{16\pi^{3}T^{2}}
          \left( \Delta^{*}\partial_{t}\Delta - \Delta\partial_{t}\Delta^{*}
          \right) .
\end{align}
We substitute Eq.~(\ref{eq:gK-expression})
with Eqs.~(\ref{eq:g^R-f})--(\ref{eq:bar_f^R-f})
into the self-consistency equation [Eq.~(\ref{eq:reduced-gap_eq})].
After tedious but straightforward calculations
with the assumption of $\tau T \gg 1$,
we find that the TDGL equation is expressed as
\begin{align}
      \label{eq:TDGL-pre}
 & \left(1-\frac{7\zeta(3)}{\pi^{3}\tau T}\right)
   \left[\partial_{t}\Delta + i(\mu_{+}-\mu_{-})\Delta\right]
          \nonumber \\
 &  = \frac{8T}{\pi}\left(1-\frac{\pi}{8\tau T}\right)
      \left( 1 - \frac{T}{T_{\rm c}} \right)\Delta
      + \frac{2T}{\pi\lambda} \zeta(x)
          \nonumber \\
 & \hspace{0mm}
    + \frac{7\zeta(3)}{2\pi^{3}T}
      \biggl\{
          \left[v_{\rm F}\partial_{x}-i(\mu_{+}+\mu_{-})
          \right]^{2}\Delta
          \nonumber \\
 & \hspace{15mm}
        + \left[\partial_{t}+i(\mu_{+}-\mu_{-})
          \right]^{2}\Delta
        - 2\Delta|\Delta|^{2}
      \biggr\}
          \nonumber \\
 & \hspace{0mm}
    + i \frac{7\zeta(3)}{2\pi^{3}T}
      \biggl\{ 
          v_{\rm F}\partial_{x}(\mu_{+}+\mu_{-})
        + \partial_{t}(\mu_{+}-\mu_{-})
        + 2v_{\rm F}eE
      \biggr\} \Delta .
\end{align}
The second term in the right-hand side represents
the impurity pinning of the CDW.

Let us return to the kinetic equation for $\hat{g}^{K}$ [Eq.~(\ref{eq:kin-G})].
Taking the trace of this equation with the help of
Eqs.~(\ref{eq:self-E_X}) and (\ref{eq:self-E_K})
and then integrating it over $\epsilon$,
we obtain the following equation:
\begin{align}
      \label{eq:continuity-pre}
 &  \int d\epsilon \,{\rm tr}
    \biggl\{
    \hat{\sigma}_{z}\partial_{t}\hat{g}^{K}
  + v_{\rm F}\partial_{x}\hat{g}^{K}
  + \left(v_{\rm F}e\dot{A}-\dot{\hat{\Delta}}+\hat{\sigma}_{z}\dot{\Phi}
    \right) \partial_{\epsilon}\hat{g}^{K}
    \biggr\}
       \nonumber \\
 & \hspace{10mm}
   = 0 .
\end{align}
Using Eqs.~(\ref{eq:charge_density}) and (\ref{eq:current_density})
with $\int d\epsilon \partial_{\epsilon}\hat{g}^{K}=4\hat{\sigma}_{z}$,
we can show that Eq.~(\ref{eq:continuity-pre}) is equivalent to
the continuity equation~\cite{eckern1}
\begin{align}
  \partial_{t}\rho + \partial_{x}j = 0 .
\end{align}
In a similar manner, we take the trace of the kinetic equation
multiplied by $\hat{\sigma}_{z}$ with the help of Eqs.~(\ref{eq:self-E_X})
and (\ref{eq:self-E_K}) and then integrate it over $\epsilon$.
This leads to
\begin{align}
 &  \int d\epsilon
    \,{\rm tr}\biggl\{
    \partial_{t}\hat{g}^{K}
  + v_{\rm F}\hat{\sigma}_{z}\partial_{x}\hat{g}^{K}
  - i\left( \hat{\sigma}_{z}\hat{\Delta}
                 -\hat{\Delta}\hat{\sigma}_{z}\right)\hat{g}^{K}
         \nonumber \\
 & \hspace{5mm}
  + \left(v_{\rm F}e\dot{A}\hat{\sigma}_{z}+\dot{\Phi}\right)
    \partial_{\epsilon}\hat{g}^{K}
  + \frac{i}{8}\left( \hat{\sigma}_{z}\ddot{\hat{\Delta}}
                         -\ddot{\hat{\Delta}}\hat{\sigma}_{z}\right)
               \partial_{\epsilon}^{2}\hat{g}^{K}
         \nonumber \\
 & \hspace{5mm}
  - \frac{i}{2}\nu_{2} 
    \left(\hat{g}^{R}-\hat{g}^{A}\right)
    \left(\hat{\sigma}_{x}\hat{g}^{K}\hat{\sigma}_{y}
              -\hat{\sigma}_{y}\hat{g}^{K}\hat{\sigma}_{x}\right)
    \biggr\} = 0 .
\end{align}
Using Eqs.~(\ref{eq:charge_density}) and (\ref{eq:current_density})
with $\int d\epsilon \partial_{\epsilon}\hat{g}^{K}=4\hat{\sigma}_{z}$
and $\int d\epsilon \partial_{\epsilon}^{2}\hat{g}^{K}=0$,
we can show that the above equation gives rise to
the BT equation:
\begin{align}
     \label{eq:Boltzmann}
 & \frac{\pi}{e}\left(\partial_{t}j + v_{\rm F}^{2}\partial_{x}\rho\right)
   - v_{\rm F}eE
         \nonumber \\
 & \hspace{5mm}
  = \frac{1}{2\tau_{\rm tr}}
    \left( 1 - \frac{7\zeta(3)}{4\pi^{2}T^{2}}|\Delta|^{2}
    \right)^{2} (\mu_{+}-\mu_{-})
    + \eta ,
\end{align}
where $\tau_{\rm tr}$ is defined in Eq.~(\ref{eq:def-tau_tr}) and
\begin{align}
       \label{eq:def-eta}
 \eta =  \frac{i}{8} \int d\epsilon \,{\rm tr}
         \left\{
            \left( \hat{\sigma}_{z}\hat{\Delta}
                   -\hat{\Delta}\hat{\sigma}_{z}\right)\hat{g}^{K}
         \right\} .
\end{align}
The first term in the right-hand side of Eq.~(\ref{eq:Boltzmann})
describes the relaxation of quasiparticles due to impurity scattering.
The expression for $\eta$ is determined
by using Eq.~(\ref{eq:reduced-gap_eq}) as
\begin{align}
      \label{eq:res-eta}
   \eta
  & = \frac{i}{4\lambda} \left(\zeta^{*}\Delta - \zeta\Delta^{*}\right)
       \nonumber \\
  & = -\frac{1}{2\lambda}V_{\rm imp}(x)|\Delta(x,t)|
       \sin\left(Qx+\theta(x,t)\right)
\end{align}
with
\begin{align}
   \theta(x,t) = {\rm arg}\{\Delta(x,t)\} .
\end{align}
This clearly indicates that $\eta$ describes the relaxation of the CDW motion
due to impurity pinning.

In the remainder of this section, we briefly consider the effective pinning
potential giving rise to the pinning term in the TDGL equation.
As the TDGL equation should be expressed in the form of
\begin{align}
   \frac{\pi}{8v_{\rm F}T}\left(1-\frac{7\zeta(3)}{\pi^{3}\tau T}\right)
   \partial_{t}\Delta
  = -\frac{\delta F(\Delta,\Delta^{*})}{\delta \Delta^{*}}
\end{align}
in terms of the free energy, written as
\begin{align}
 & F = \int dx \frac{1}{v_{\rm F}}
       \biggl[ - \left(1-\frac{\pi}{8\tau T}\right)
                \left( 1-\frac{T}{T_{\rm c}} \right)
                \Delta\Delta^{*}
         \nonumber \\
 & \hspace{20mm}
              + \frac{7\zeta(3)}{16\pi^{2}T^{2}}(\Delta\Delta^{*})^{2}
              + \cdots \biggr] + F_{\rm pin} ,
\end{align}
the pinning potential $F_{\rm pin}$ is identified as
\begin{align}
  F_{\rm pin}
  & = - \frac{1}{4\lambda v_{\rm F}} \int dx
        \left( \zeta^{*}\Delta+\zeta\Delta^{*} \right)
             \nonumber \\
  & = - \frac{1}{2\lambda v_{\rm F}} \int dx V_{\rm imp}(x)|\Delta(x)|
        \cos\left(Qx+\theta(x)\right) .
\end{align}
The $\lambda$ dependence of $F_{\rm pin}$
is consistent with Eq.~(6.9) of Ref.~\citen{tutto}.
With the expression for $V_{\rm imp}$ [Eq.~(\ref{eq:def-V_imp})],
the pinning potential is rewritten in
the following well-known form:~\cite{fukuyama1,fukuyama2}
\begin{align}
  F_{\rm pin}
    = - \frac{1}{2\lambda v_{\rm F}} \int dx  \sum_{i} v(x-x_{i})|\Delta(x)|
        \cos\left(Qx+\theta(x)\right) .
\end{align}

\section{Application of the TDGL and BT Equations}

The TDGL equation [Eq.~(\ref{eq:TDGL-pre})] with Eq.~(\ref{eq:def-zeta})
and the BT equation [Eq.~(\ref{eq:Boltzmann})] with Eq.~(\ref{eq:res-eta})
are the central result of this paper.
Although the effect of the Coulomb interaction is not explicitly considered
in their derivation, we can take account of it including the screening effect
due to quasiparticles by supplementing them
by Gauss's law.~\cite{ hayashi1,hayashi2}
The most interesting application of the TDGL and BT equations is to use them
for numerical simulations of the nonequilibrium dynamics
of the CDW and quasiparticles in phase slip processes.~\cite{hayashi2}
Such a numerical study will be reported in a forthcoming publication.
Here, by using the two equations, we analyze the dc electric conductivity
of CDW conductors in the situation where a space- and time-independent
current density (i.e., $j=$ constant) is supplied to the system.
The factor $(\tau T)^{-1}$ in the TDGL equation is ignored
in the following argument as it does not play an important role.

Firstly, we consider the case where the electric field $E$
is much smaller than the threshold value $E_{\rm th}$ for CDW depinning
and hence the CDW is completely pinned.
That is, $\theta(x,t)$ as well as $|\Delta(x,t)|$ is independent of $t$
but varies as a function of $x$ according to the pinning potential.
As only quasiparticles contribute to the current density
in this case, the chemical potentials for right-going and left-going
quasiparticles should satisfy $\mu_{+}-\mu_{-} =$ constant.
The TDGL equation reads
\begin{align}
     \label{eq:TDGL-static}
 & i\left(\mu_{+}-\mu_{-}\right)|\Delta|
         \nonumber \\
 & = \frac{8T}{\pi}\left(1-\frac{T}{T_{\rm c}} \right)|\Delta|
     + \frac{2T}{\pi\lambda}|\zeta|e^{-i(Qx+\theta)}
         \nonumber \\
 & \hspace{2mm}
     + \frac{7\zeta(3)}{2\pi^{3}T}
       \Bigl[  v_{\rm F}^{2}
                 \left(  \partial_{x}^{2}|\Delta|
                       -(\partial_{x}\theta)^{2}|\Delta| \right)
         \nonumber \\
 & \hspace{8mm}
               + 2v_{\rm F}(\mu_{+}+\mu_{-})(\partial_{x}\theta)|\Delta|
               - 2(\mu_{+}^{2}+\mu_{-}^{2})|\Delta| - 2|\Delta|^{3}
       \Bigr]
         \nonumber \\
 & \hspace{2mm}
     + i \frac{7\zeta(3)}{2\pi^{3}T}
       \Bigl[ v_{\rm F}^{2}
               \left( (\partial_{x}^{2}\theta)|\Delta|
                      +2(\partial_{x}\theta)(\partial_{x}|\Delta|) \right)
         \nonumber \\
 & \hspace{8mm}
               - 2v_{\rm F}(\mu_{+}+\mu_{-})\partial_{x}|\Delta|
               + 2v_{\rm F}eE|\Delta|
       \Bigr]
\end{align}
and the BT equation is expressed as
\begin{align}
      \label{eq:BT-case1}
 & - \frac{v_{\rm F}}{2}
     \left(1-\frac{7\zeta(3)}{4\pi^{2}T^{2}}|\Delta|^{2}\right)
     \partial_{x}\left(\mu_{+}+\mu_{-}\right)
          \nonumber \\
 & - \frac{7\zeta(3)}{4\pi^{2}T^{2}}v_{\rm F}
     \biggl[  \frac{v_{\rm F}}{2}(\partial_{x}^{2}\theta)|\Delta|^{2}
            + v_{\rm F} (\partial_{x}\theta)(\partial_{x}|\Delta|)|\Delta|
          \nonumber \\
 & \hspace{23mm}
            -(\mu_{+}+\mu_{-})(\partial_{x}|\Delta|)|\Delta|
     \biggr] - v_{\rm F}eE
          \nonumber \\
 & \hspace{-2mm}
   = \frac{1}{2\tau_{\rm tr}}
     \left( 1 - \frac{7\zeta(3)}{4\pi^{2}T^{2}}|\Delta|^{2}
     \right)^{2} (\mu_{+}-\mu_{-})
          \nonumber \\
 & - \frac{1}{2\lambda}|\zeta||\Delta|
     \sin\left(Qx+\theta\right).
\end{align}
We take $\partial_{x}\left(\mu_{+}+\mu_{-}\right)$ into consideration
in Eq.~(\ref{eq:BT-case1}) since the spatial variation of $\theta(x,t)$
leads to an inhomogeneity in the charge density of the CDW,
which then induces a spatial variation of the quasiparticle charge density
determined by $\mu_{+}+\mu_{-}$.
This phenomenon (i.e., the screening of the charge density due to
quasiparticles) is described by Gauss's law.
However, its details including the resulting profile of $\Delta$
do not affect the argument given below.
The imaginary part of Eq.~(\ref{eq:TDGL-static}) yields
\begin{align}
 &  \left(\mu_{+}-\mu_{-}\right)|\Delta|
            \nonumber \\
 & \hspace{-2mm}
   = \frac{7\zeta(3)}{4\pi^{3}T}v_{\rm F}
     \biggl[ \frac{v_{\rm F}}{2}\left( (\partial_{x}^{2}\theta)|\Delta|
                           + (\partial_{x}\theta)(\partial_{x}|\Delta|)
                        \right)
            \nonumber \\
 & \hspace{5mm}
             - (\mu_{+}+\mu_{-})\partial_{x}|\Delta| + eE|\Delta|
     \biggr]
  - \frac{2T}{\pi\lambda}|\zeta|\sin\left(Qx+\theta\right). 
\end{align}
The combination of the last two equations yields
\begin{align}
 & - v_{\rm F}e\tilde{E}
 = \frac{1}{2\tau_{\rm tr}}
     \left( 1 - \frac{7\zeta(3)}{4\pi^{2}T^{2}}|\Delta|^{2}\right)
     \left(1+\frac{\pi\tau_{\rm tr}|\Delta|^{2}}{2T}\right)
           \nonumber \\
 & \hspace{40mm}
     \times
     \left(\mu_{+}-\mu_{-}\right) ,
\end{align}
where
\begin{align}
  \tilde{E} = E + \frac{1}{2e}\partial_{x}\left(\mu_{+}+\mu_{-}\right) .
\end{align}
Note that $\tilde{E}$ corresponds to
the derivative of the electrochemical potential.
By using this relation, we eliminate $\mu_{+}-\mu_{-}$
in the expression for the current density, which reads
\begin{align}
        \label{eq:j-case1}
  j = -\frac{e}{2\pi}
       \left(1-\frac{7\zeta(3)}{4\pi^{2}T^{2}}|\Delta|^{2} \right)
       (\mu_{+}-\mu_{-})
\end{align}
in this case.
We finally find that $j = \sigma \tilde{E}$
with the conductivity $\sigma$ given by
\begin{align}
     \label{eq:result1}
  \sigma
  = \sigma_{\rm N}\left(1-\frac{\pi\tau_{\rm tr}|\Delta|^{2}}{2T}\right) ,
\end{align}
where $\sigma_{\rm N} \equiv e^{2}v_{\rm F}\tau_{\rm tr}/\pi$
is the conductivity in the normal state.
We observe that the conductivity is smaller than the normal-state value.
This should not simply be attributed to
the suppression of the density of states, which certainly decreases $j$
as shown in Eq.~(\ref{eq:j-case1}) but effectively
increases $\tau_{\rm tr}$ according to Eq.~(\ref{eq:BT-case1}).
As the factor of the increase in $\tau_{\rm tr}$ is larger than
the factor of the decrease in $j$, the conductivity is enhanced,
contrary to the above result,
if only these two changes are taken into consideration.
The reduction of $\sigma$ below the normal-state value is mainly caused by
the screening of the electric field due to the CDW,
which dominates the effect of the suppression of the density of states.

Secondly, we consider the case where $E \gg E_{\rm th}$
and hence the CDW moves in the direction of acceleration.
In this case, the CDW motion can be approximated as uniform sliding
as long as our attention is focused on the dc conductivity.
That is, $|\Delta(x,t)| = {\rm constant}$ and the phase increases with
the angular velocity $\omega$ as $\theta(x,t) = -\omega t +{\rm constant}$.
In accordance with this approximation, we assume that $\mu_{+}+\mu_{-} = 0$,
resulting in $\rho = 0$.
Note that the current density arises from both the CDW and quasiparticles
in this case.
The TDGL equation reads
\begin{align}
     \label{eq:TDGL-dynamic}
 & i\left(-\omega+\mu_{+}-\mu_{-}\right)|\Delta|
         \nonumber \\
 & = \frac{8T}{\pi}\left(1-\frac{T}{T_{\rm c}} \right)|\Delta|
     + \frac{7\zeta(3)}{2\pi^{3}T}
       \Bigl[- \left(-\omega+\mu_{+}-\mu_{-}\right)^{2}
      \nonumber \\
 & \hspace{40mm}
             - 2|\Delta|^{2} + i2 v_{\rm F}eE
       \Bigr]|\Delta| ,
\end{align}
and the BT equation is expressed as
\begin{align}
       \label{eq:Boltzmann-dynamic}
   - v_{\rm F}eE
 & = \frac{1}{2\tau_{\rm tr}}
     \left( 1 - \frac{7\zeta(3)}{4\pi^{2}T^{2}}|\Delta|^{2}
     \right)^{2} (\mu_{+}-\mu_{-}) ,
\end{align}
where the pinning term, which oscillates with the frequency $\omega$,
is ignored in each equation
since our attention is focused on the dc conductivity.
The imaginary part of Eq.~(\ref{eq:TDGL-dynamic}) yields
\begin{align}
     \label{eq:TDGL1}
  -\omega + \mu_{+}-\mu_{-}
 = \frac{7\zeta(3)}{\pi^{3}T}v_{\rm F}eE .
\end{align}
Now, we derive the dc conductivity from the expression for the current density:
\begin{align}
      \label{eq:current1}
  j = -\frac{e}{2\pi}
       \left(1-\frac{7\zeta(3)}{4\pi^{2}T^{2}}|\Delta|^{2} \right)
       (\mu_{+}-\mu_{-})
      -e\frac{7\zeta(3)}{8\pi^{3}T^{2}}\omega|\Delta|^{2} ,
\end{align}
where the second term corresponds to the CDW current.
Using Eqs.~(\ref{eq:Boltzmann-dynamic}) and (\ref{eq:TDGL1}), we can rewrite
$\mu_{+}-\mu_{-}$ and $\omega$ in Eq.~(\ref{eq:current1})
in terms of $E$, and we find that $j = \sigma E$ with
\begin{align}
     \label{eq:result2}
  \sigma
  = \sigma_{\rm N}
    \left[ 1 + \frac{7\zeta(3)|\Delta|^{2}}{2\pi^{2}T^{2}}
                \left(1+\frac{7\zeta(3)}{4\pi^{3}\tau_{\rm tr}T}\right)
    \right] .
\end{align} 
The conductivity is slightly larger than the normal-state value.~\cite{gorkov3}
This is mainly caused by the increase in $\tau_{\rm tr}$ due to
the suppression of the density of states without the decrease in $j$,
which is compensated by the additional contribution from the sliding CDW.

We see that $\sigma$ is smaller than $\sigma_{\rm N}$ in the regime of
$E \ll E_{\rm th}$ while it is slightly larger than $\sigma_{\rm N}$
in the opposite regime of $E \gg E_{\rm th}$.
This clearly indicates the nonlinear behavior of electric conductivity
in CDW conductors.
It should be mentioned that Eq.~(\ref{eq:result1}) is equivalent to
Eq.~(34) of Ref.~\citen{artemenko1} and that Eq.~(\ref{eq:result2}) is also
equivalent to the equation presented
just below Eq.~(35) of Ref.~\citen{artemenko1}.
Our derivation is much simpler than that in Ref.~\citen{artemenko1},
which was based on the Keldysh Green's function approach
under a quasiclassical approximation.
Furthermore, the effect of impurity pinning was treated
in a phenomenological manner in Ref.~\citen{artemenko1},
while we explicitly take it into consideration.
In these respects, our theoretical framework has an advantage over
the previous one.

\section{Summary}

We have derived the time-dependent Ginzburg--Landau equation
and the Boltzmann transport equation for charge-density-wave (CDW) conductors
from a microscopic one-dimensional model
by applying the Keldysh Green's function approach
under a quasiclassical approximation.
We have succeeded in introducing the pinning term
without relying on a phenomenological argument.
These equations simultaneously describe the spatiotemporal dynamics
of both the CDW and quasiparticles;
thus, they can be widely used to analyze various nonequilibrium phenomena
that are associated with both their degrees of freedom.
For example, they enable us to numerically simulate the dynamics of
the CDW and quasiparticles in phase slip processes including
the effects of an external electric field and impurity pinning.~\cite{hayashi2}
An extension of this framework to two- and three-dimensional cases
is the most important achievement to be accomplished.

\section*{Acknowledgment}

This work was partially supported by JSPS KAKENHI Grant Numbers
15K05130 and JP24540392.

\appendix

\section{Derivation of the Self-Consistency Equation}

In this appendix, we derive the self-consistency equation for $\Delta(x,t)$
following the argument in Ref.~\citen{artemenko1}.
Let us consider the Hamiltonian
$H^{\prime} = H_{\rm ph}+H_{\rm e\mathchar`-p}$,
which describes only phonon degrees of freedom
including the electron-phonon coupling,
where $H_{\rm ph}$ and $H_{\rm e\mathchar`-p}$ are given as follows:
\begin{align}
 H_{\rm ph}
   & =   \sum_{q} \omega_{q}
         \left(b_{q}^{\dagger}b_{q}+\frac{1}{2}\right) ,
       \\
 H_{\rm e\mathchar`-p}
   & =   \int dx
         \Big\{ g u(x)e^{-iQx}\psi_{+}^{\dagger}(x)\psi_{-}(x)
                + {\rm h.c.}
         \Big\} .
\end{align}
The lattice displacement $u(x)$ is expressed
in terms of the phonon operators as
\begin{align}
  u(x) = \frac{1}{\sqrt{L}}\sum_{q}\frac{1}{\sqrt{2\rho_{\rm ph}\omega_{q}}}
         \left(b_{q}+b_{-q}^{\dagger}\right) e^{-iqx} ,
\end{align}
where $L$ is the system length and $\rho_{\rm ph}$ is the mass density.
With the Fourier transform of $\psi_{\pm}(x)$:
\begin{align}
   \psi_{\pm}(k) = \frac{1}{\sqrt{L}}\int dx e^{-ikx}\psi_{\pm}(x) ,
\end{align}
it is convenient to rewrite $H_{\rm e\mathchar`-p}$ as
\begin{align}
 H_{\rm e\mathchar`-p}
   & =   \frac{1}{\sqrt{L}}\sum_{k,q}
         \frac{g}{\sqrt{2\rho_{\rm ph}\omega_{Q+q}}}
         \left(b_{Q+q}+b_{-(Q+q)}^{\dagger}\right)
                  \nonumber \\
   & \hspace{3mm}
         \times
         \left\{ \psi_{+}^{\dagger}(k_{+})\psi_{-}(k_{-})
               + \psi_{-}^{\dagger}(k_{-})\psi_{+}(k_{+})
         \right\} ,
\end{align}
where $k_{\pm} = k \pm q/2$.
From the Heisenberg equation for the phonon operators,
we can show that
\begin{align}
  & \left(\omega_{Q}^{2}+\partial_{t}^{2}\right)
    \left\langle b_{Q+q}(t)+b_{-(Q+q)}^{\dagger}(t)\right\rangle
       \nonumber \\
  & \hspace{0mm}
    = \frac{1}{\sqrt{L}}\frac{g(-2\omega_{Q})}{\sqrt{2\rho_{\rm ph}\omega_{Q}}}
      \sum_{k} \left\langle
                  \psi_{-}^{\dagger}(k_{-},t)\psi_{+}(k_{+},t)
               \right\rangle .
\end{align}
Performing the inverse Fourier transformation of the above equation
and using the relation $g\langle u \rangle e^{-iQx} = \Delta - \zeta$,
we obtain
\begin{align}
 & \left(1+\omega_{Q}^{-2}\partial_{t}^{2}\right)
   \left[\Delta(x,t)-\zeta(x)\right]
          \nonumber \\
 & \hspace{0mm}
   = i\frac{g^{2}}{2\rho_{\rm ph}\omega_{Q}^{2}}
     \left. G_{+-}^{K}(x,t;x',t')\right|_{x'\to x,t'\to t} .
\end{align}
Ignoring the irrelevant term with $\partial_{t}^{2}$
and rewriting the prefactor in the right-hand side in terms of
the dimensionless coupling constant
\begin{align}
  \lambda = \frac{g^{2}}{8\pi\rho_{\rm ph}\omega_{Q}^{2}v_{\rm F}} ,
\end{align}
we finally arrive at the self-consistency equation [Eq.~(\ref{eq:gap_eq0})].

\section{Decomposition of Eq.~(\ref{eq:kin-G})}

We decompose Eq.~(\ref{eq:kin-G}) to yield a set of four equations
for the matrix elements of $\hat{g}^{R}$.
The resulting equations are as follows:
\begin{align}
 & i\left(\partial_{t}+v_{\rm F}\partial_{x}\right)g^{R}
   + \Delta^{*} f^{R} + \Delta \bar{f}^{R}
   + i\left( ev_{\rm F}\dot{A} + \dot{\Phi} \right)
            \partial_{\epsilon}g^{R}
        \nonumber \\
 & \hspace{0mm}
   + \frac{i}{2}\left(\dot{\Delta}^{*}\partial_{\epsilon}f^{R}
                    - \dot{\Delta}\partial_{\epsilon}\bar{f}^{R}
                \right)
   - \frac{1}{8}\left(\ddot{\Delta}^{*}\partial_{\epsilon}^{2}f^{R}
                      + \ddot{\Delta}\partial_{\epsilon}^{2}\bar{f}^{R}
                \right)
 = 0 ,
        \\
 & i\left(-\partial_{t}+v_{\rm F}\partial_{x}\right)\bar{g}^{R}
   - \Delta^{*} f^{R} - \Delta \bar{f}^{R}
   + i\left(ev_{\rm F}\dot{A}-\dot{\Phi} \right)
            \partial_{\epsilon}\bar{g}^{R}
        \nonumber \\
 & \hspace{0mm}
   + \frac{i}{2}\left(\dot{\Delta}^{*}\partial_{\epsilon}f^{R}
                    - \dot{\Delta}\partial_{\epsilon}\bar{f}^{R}
                \right)
   + \frac{1}{8}\left(\ddot{\Delta}^{*}\partial_{\epsilon}^{2}f^{R}
                      + \ddot{\Delta}\partial_{\epsilon}^{2}\bar{f}^{R}
                       \right)
 = 0 ,
        \\
 & 2(\epsilon-\Phi)f^{R}
   + iv_{\rm F}\partial_{x}f^{R}-\Delta(g^{R}-\bar{g}^{R})
   + iev_{\rm F}\dot{A}\partial_{\epsilon}f^{R}
        \nonumber \\
 & \hspace{0mm}
   - \frac{i}{2}\dot{\Delta}
     \left(\partial_{\epsilon}g^{R} + \partial_{\epsilon}\bar{g}^{R}
     \right)
   + \frac{1}{8}\ddot{\Delta}
     \left(\partial_{\epsilon}^{2}g^{R}
           - \partial_{\epsilon}^{2}\bar{g}^{R} \right)
        \nonumber \\
 & \hspace{0mm}
   + \frac{i}{2\tau}(g^{R}-\bar{g}^{R})f^{R} = 0,
        \\
 & - 2(\epsilon-\Phi)\bar{f}^{R} + iv_{\rm F}\partial_{x}\bar{f}^{R}
       - \Delta^{*}(g^{R}-\bar{g}^{R})
   + iev_{\rm F}\dot{A}\partial_{\epsilon}\bar{f}^{R}
        \nonumber \\
 & \hspace{0mm}
   + \frac{i}{2}\dot{\Delta}^{*}
     \left(\partial_{\epsilon}g^{R} + \partial_{\epsilon}\bar{g}^{R}
     \right)
   + \frac{1}{8}\ddot{\Delta}^{*}
     \left(\partial_{\epsilon}^{2}g^{R}
           - \partial_{\epsilon}^{2}\bar{g}^{R} \right)
        \nonumber \\
 & \hspace{0mm}
   - \frac{i}{2\tau}(g^{R}-\bar{g}^{R})\bar{f}^{R} = 0 .
\end{align}


\begin{thebibliography}{99}
\bibitem{comment} At this stage, it is possible for $\Delta(x,t)$ to contain
components that spatially vary on the scale of the Fermi wavelength
owing to the presence of $\zeta(x)$ in its definition.
However, such components should be subtracted in the resulting TDGL equation
as it describes the dynamics of $\Delta(x,t)$ on a much longer length scale.
\end{thebibliography}
\end{document}